\documentclass[reprint, showpacs, aps, nofootinbib, superscriptaddress, prb]{revtex4-1}
\usepackage{amsmath}   
\usepackage{graphicx}
\usepackage{epstopdf}
\usepackage{hyperref}
\usepackage[utf8]{inputenc}
\usepackage[english]{babel}

\usepackage{soul}
\usepackage[usenames,dvipsnames]{color}
\usepackage[normalem]{ulem}

\begin{document}

\title{$^{17}$O--NMR Knight shift study of the interplay between superconductivity and pseudogap in (Ca$_x$La$_{1-x}$)(Ba$_{1.75 - x}$La$_{0.25 + x}$)Cu$_3$O$_y$}

\author{T. Cvitani\'{c}}
\affiliation{Department of Physics, Faculty of Science, University of Zagreb, Bijeni\v{c}ka 32, HR-10000, Zagreb, Croatia}
\author{D. Pelc}
\affiliation{Department of Physics, Faculty of Science, University of Zagreb, Bijeni\v{c}ka 32, HR-10000, Zagreb, Croatia}
\author{M. Po\v{z}ek}
\affiliation{Department of Physics, Faculty of Science, University of Zagreb, Bijeni\v{c}ka 32, HR-10000, Zagreb, Croatia}
\author{E. Amit}
\affiliation{Department of Physics, Technion -- Israel Institue of Technology, Haifa, 32000, Israel}
\author{A. Keren}
\affiliation{Department of Physics, Technion -- Israel Institue of Technology, Haifa, 32000, Israel}

\begin{abstract}

We report systematic $^{17}$O--NMR measurements on the high-$T_c$ cuprate (Ca$_x$La$_{1-x}$)(Ba$_{1.75 - x}$La$_{0.25 + x}$)Cu$_3$O$_y$, for four different families (different $x$). Using Knight shift data, we show that the pseudogap opening temperature $T*$ is much higher than $T_c$ near optimal doping, unlike structurally similar YBCO. In addition, at constant doping the pseudogap temperature does not vary with $x$, in contrast to $T_c$. This puts constraints on the nature of the pseudogap and position of the quantum critical point inside the superconducting dome.

\end{abstract}


\maketitle

\section{INTRODUCTION}
The pseudogap (PG) is still one of the most important and yet among the least understood features of cuprate physics. \cite{Norman03,Norman_Pines_03,Lee_06,Hufner_2008,Rice_12} It is experimentally seen as a partial gapping of the electronic spectrum below an onset temperature $T^*$, mostly in the underdoped part of the cuprate phase diagram. Although such a gapped area of the phase diagram has been found in all hole-doped high-$T_c$ superconductors so far, the universality of its characteristics is controversial. In the underdoped region, the pseudogap onset temperature $T^*$ decreases as hole doping increases, but in the optimally doped and overdoped regions the behaviour (and indeed the existence) of the pseudogap is uncertain. It has been suggested that the PG line intersects the superconducting (SC) dome in the phase diagram -- as seems to be the case in  the most studied cuprate YBa$_2$Cu$_3$O$_y$ (YBCO) -- or merges with the dome on the overdoped side. \cite{Alloul_10} Although it has long been speculated that the pseudogap is in some way related to high-$T_c$ superconductivity, these conflicting possibilities preclude any general agreement on the nature of such a relation. The pseudogap state could be a direct precursor to superconductivity, coexist independently, or compete. \cite{Norman_Pines_03,Millis_06,Cho06,Kondo_09} It is even ambiguous whether $T^*$ is a true phase transition \cite{Shekhter_13} or simply a crossover temperature.

Several theories have been proposed to
explain the phase diagram, including the loop current
model,\cite{Varma_99} orbital density wave (ODW),\cite{Nayak_00} or interpenetrating
spin-orbital density waves.\cite{Laughlin_14}
An idea which has recently attracted increased attention due to results on YBCO is a connection between superconductivity, pseudogap and quantum critical behaviour. \cite{Shekhter_13} In that picture the pseudogap line is a line of real phase transitions, intersecting the superconducting dome close to optimal doping and ending in a quantum critical point \cite{Sondhi_97} (QCP) at $T = 0$. A loop current model has been proposed in this context to explain the phase diagram. \cite{Varma_99} Several experiments indicate the existence of circular currents in underdoped YBCO, \cite{Fauque_06,Kapitulnik_09} and recently evidence was found that a thermodynamic phase transition indeed occurs at (or close to) $T^*$.  \cite{Shekhter_13} These experiments bring up the important question whether such a model for the pseudogap is universal in all cuprates, and, more generally, if high-$T_c$ superconductivity as such is a consequence of quantum criticality? Due to several advantages -- relative structural homogeneity, high transition temperatures, and quality of available crystals -- YBCO is the most studied cuprate \cite{Nagasao_08,Fradkin_12} (at least in terms of pseudogap physics) and there is a tendency to regard results on YBCO as universal. It is our purpose to discuss pseudogap data for a structurally very similar system of cuprates -- (Ca$_x$La$_{1-x}$)(Ba$_{1.75 - x}$La$_{0.25 + x}$)Cu$_3$O$_y$ (CLBLCO), in order to check PG universality in cuprates. We show strong experimental evidence that the quantum critical behaviour of the pseudogap is not universal, as $T^*$ is still very far from SC dome in optimally doped CLBLCO, in spite of very similar critical temperatures and structural features compared to YBCO. Our data instead indicate that pseudogap and superconductivity are coexisting phenomena, with no simple relation between them.

\section{SAMPLES AND EXPERIMENTAL TECHNIQUE}
\begin{figure}
\includegraphics[width = 60mm]{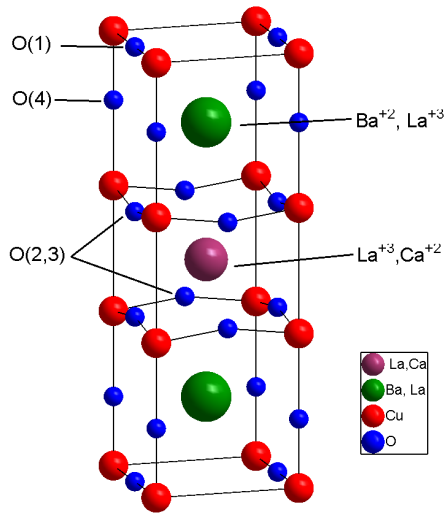}
\caption{(Color online) CLBLCO unit cell, which is tetragonal and very similar to YBCO cuprate. There is no ordering of chain oxygen O(1).}
\label{fig:struct}
\end{figure}

(Ca$_x$La$_{1-x}$)(Ba$_{1.75 - x}$La$_{0.25 + x}$)Cu$_3$O$_y$ is a unique system in which one may chemically control not only the oxygen doping $y$, but also the maximum $T_c$ for the optimal oxygen doping (by changing $x$). It has an YBCO-like structure for all families ($x$) and oxygen doping levels ($y$) \cite{Chmaissem_01} (see Fig.~\ref{fig:struct}). All samples are tetragonal and there is no chain ordering. CLBLCO families have negligible structural differences, \cite{Goldschmidt_93} but the highest superconducting transition temperature varies up to 30\% between families. \cite{Ofer06} $T_c$ is around 80~K for optimally doped ($y \approx 7.15$) sample from the $x = 0.4$ family -- close to optimally doped YBCO. One can roughly imagine the CLBLCO families to be an extension of YBCO, with the advantageous possibility of systematically tuning electronic interactions by changing the family $x$ without drastic changes in crystal structure. Similar structures with varying superconducting and magnetic properties thus make CLBLCO ideal for understanding the relation between superconductivity and pseudogap temperature.

The level of disorder in the CLBLCO system has been investigated by several experimental techniques. The most direct measurement was done by High resolution powder X-ray diffraction (HRPXRD) studies. \cite{Agrestini_14} It was found that line widths of the (006) and (200) Bragg peaks and the isotropic atomic displacement factor for the Ba/La site, for the optimally doped CLBLCO, slightly increase as $x$ increases. This indicates disorder increase, mainly on the Ba site, as more Ca is introduced into the system. This anticipated result cannot explain the increase in $T_c$ as $x$ increases. A clearer result was obtained by Raman scattering which shows that the most pronounced phonon peak at 300~cm$^{-1}$ is in fact narrower for the $x=0.1$ than for the $x=0.4$ family. The width of this peak is a measure of the phonon coherence length. The cleaner the material the longer is this length and the narrower is the peak. \cite{Wulferding_14} The Raman scattering measurements indicate that despite its lower $T_c$, the $x=0.1$ is cleaner. Angle resolved photo emission spectroscopy (ARPES) measurements were also performed on CLBLCO. \cite{Drachuck_14} They determined the electronic band structure. While the line width of the energy dispersion curve (EDC) peak at the Fermi momentum ($k_F$) is a measure of the single particle lifetime, the line width of the momentum dispersion curve (MDC) at the Fermi energy ($E_F$) is a measure of the single particle coherence length. No quantitative differences were found between the EDC line widths. The MDC of the $x=0.1$ sample is broader in the nodal direction, and narrower in the anti-nodal direction. Since the anti-nodal direction is more relevant to HTSC, it seems that from the ARPES point of view, the $x=0.1$ sample is again less disordered one.
Finally, NMR on Cu, \cite{Keren_09} O (Ref.~[\onlinecite{Amit_10}] and our high--resolution measurements below) and Ca \cite{Marchand} show that the line width of all nuclei are the same for both families. The NMR line width is a measure of the local distortions and magnetic impurities next to the detected nuclei. Identical line width implies identical local environment for both families. All measurements point that family with higher maximum $T_c$ has more or equal disorder, which means that it is not a relevant parameter in the CLBLCO system.

\begin{figure}
\includegraphics[width=80mm]{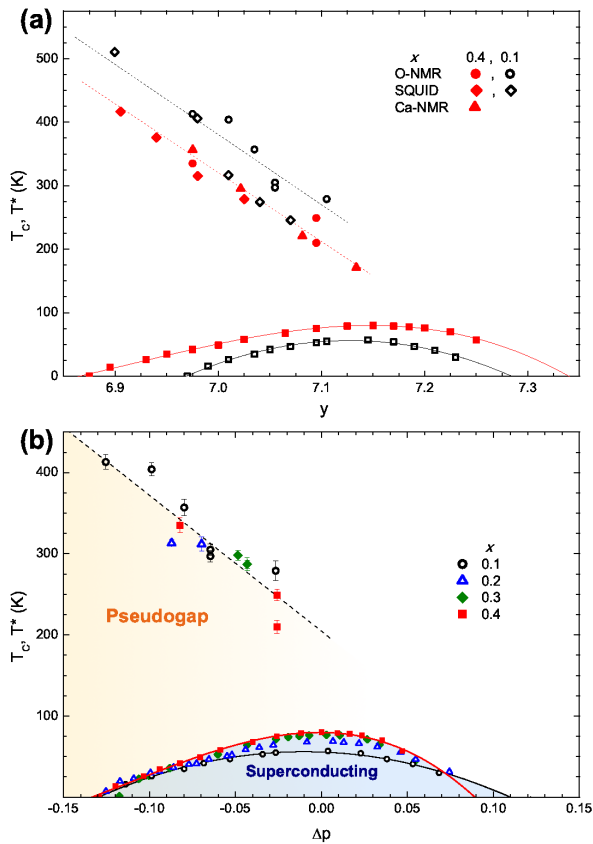}
\caption{(Color online) \textbf{(a)}  CLBLCO phase diagram. Comparison between O-NMR measurements presented in this paper and Ca-NMR and SQUID susceptibility measurements from Refs.~\onlinecite{Marchand,Lubashevsky_08}. Presented are only $x = 0.1$ and $x = 0.4$ samples. Abscissa is oxygen doping parameter $y$. Lines are guides to the eye.
\textbf{(b)} Phase diagram for all measured samples. Chemical doping $y$ is replaced with the in-plane hole concentration relative to in-plane optimal doping, $\Delta p$ (see text). The dashed pseudogap line is guide to the eye.}
\label{fig:pd}
\end{figure}

We present $^{17}$O nuclear magnetic resonance (NMR) results for four different CLBLCO families ($x = 0.1 - 0.4$). The $x=0.1$ family was comprehensively studied by measuring six samples with different oxygen doping, while the other families ($x=0.2-0.4$) were represented by two or three different dopings. For our measurements samples were prepared as described in Ref.~\onlinecite{Amit_10}. All samples were characterised with a SQUID magnetometer and all show single--component superconducting transition. Oxygen concentration was determined using well-known CLBLCO phase diagram (Fig.~\ref{fig:pd}a). \cite{Amit_10} The CLBLCO samples were in powder form, enriched with the NMR active oxygen isotope with nuclear spin $I = 5/2$. Only underdoped and nearly optimally doped samples have been measured because of technical difficulties in enriching the overdoped samples with $^{17}$O. Preparation of overdoped samples includes high pressure or liquid oxygen. Due to the price of $^{17}$O it cannot be done with this isotope.

NMR spectra were acquired with a Tecmag Apollo spectrometer, in an Oxford superconducting variable-field magnet at different magnetic fields in the vicinity of 11.5~T. Spectra were acquired by a standard Hahn echo sequence, followed by Fourier transform of the echo signal. The only previous $^{17}$O NMR study on CLBLCO \cite{Amit_10} was concerned with the nuclear quadrupole resonance parameter $\nu_Q$ -- here we focus only on the central transition ($-1/2~\leftrightarrow~1/2$) where we found two distinct NMR lines from different oxygen sites.

\section{RESULTS}
\begin{figure}
\includegraphics[width=80mm]{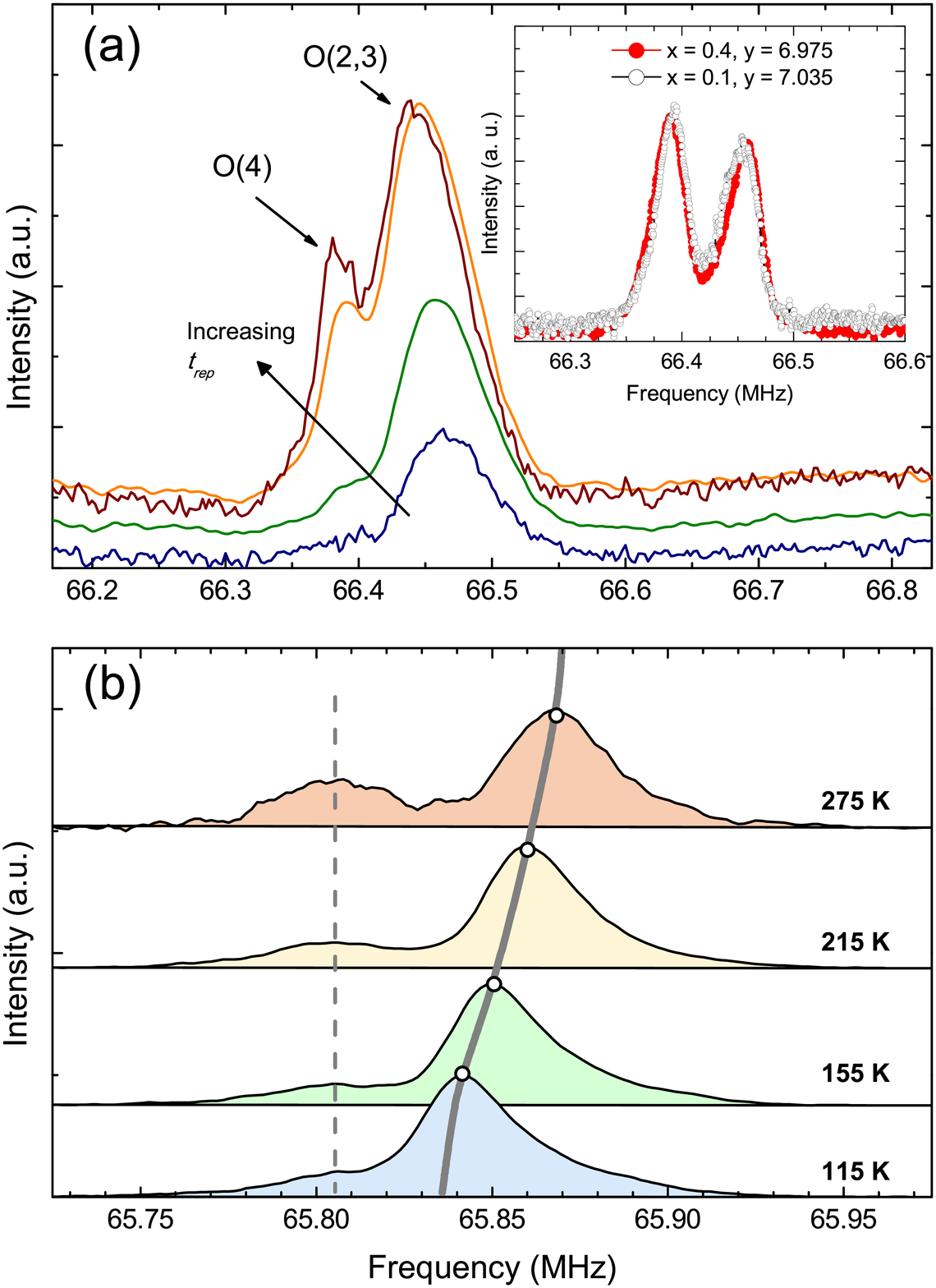}
\caption{(Color online) (a) Central NMR transition of $^{17}$O with two distinct lines. Powder spectrum recorded with repetition times ($t_{rep}$) of 3 (black), 10, 150 and 300~ms (red). The apical oxygen site O(4) has much slower spin-lattice relaxation than in-plane oxygen O(2,3). Sample: $x = 0.4$, $y = 7.1$. Inset: comparison of two different families ($x$ = 0.1 and 0.4) at (approximately) same oxygen doping $y$. The difference between linewidths are 6\% for apical and negligible (less than 1\%) for in-plane oxygen. (b) Spectra at different temperatures. Thick grey line follows in-plane oxygen peaks. The shift of the apical oxygen is temperature-independent, as is denoted by dashed line. Intensity of apical oxygen varies because its relaxation time shortens with temperature. Sample: $x = 0.1$, $y = 7.05$.}
\label{fig:spec}
\end{figure}

In Fig.~\ref{fig:spec}a we show spectrum for different repetition times ($t_{rep}$) between NMR data acquisitions. For slow nuclear spin-lattice relaxations ($t_{rep} < 5 T_1$) nuclei cannot return to equilibrium in time for another acquisition. This results in signal reduction. \cite{Slichter} By increasing repetition time we get saturation for higher frequency line at $t_{rep} > 150$~ms, while the lower frequency line remains unsaturated, implying slower spin-lattice relaxation. This is expected for cuprate superconductors. \cite{Horvatic,Oldfield_89} Based on YBCO results, we assign the line at lower frequency (slowly relaxing) to apical oxygen O(4), while the fast-relaxing line at higher frequency comes from the in-plane oxygen O(2,3). The third oxygen site O(1) which would correspond to the chain oxygen in YBCO, is difficult to see in powder samples due to disorder-induced broadening (O(1) is far from the CuO$_2$ planes where disorder is significant), and we do not observe it. 

The $^{17}$O--lines are similar in width in all measured samples, leading us to conclude that structural disorder is constant through all families (as seen by NMR). Comparison of disorder in different families can be seen in inset of Fig.~\ref{fig:spec}a. Gaussian fits to spectra give negligible linewidth difference for planar oxygen. In contrast, apical oxygen in $x = 0.4$ sample has 6\% wider line than $x = 0.1$. This confirms that disorder is small and far from CuO$_2$ planes. It is thus safe to say that any differing behaviour is purely due to electronic properties.

\begin{figure}
\includegraphics[width=80mm]{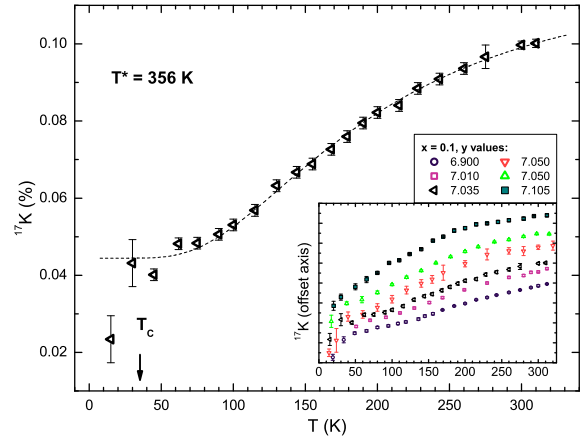}
\caption{(Color online) In-plane oxygen line position in reference to apical oxygen position for sample $x = 0.1$, $y = 7.035$. For fitting, only points above the superconducting transition $T_c$ are taken into account. As the gap opening is not abrupt, the pseudogap temperature $T^*$ is not easily distinguished in the plots, but acquired as parameter from a fit function (dotted line - see text). Inset: Knight shifts for $x = 0.1$ samples with various $y$. Each sample is offset on vertical axis for better display. Topmost sample is near-optimal doping, and its shift still strongly depends on temperature.}
\label{fig:shift}
\end{figure}

The in-plane oxygen shows a temperature-dependent Knight shift, while the apical site has no visible shift (Fig.~\ref{fig:spec}b), also in agreement with YBCO. \cite{Walstedt} Using measured values of the Knight shift, one can determine the in-plane spin susceptibility at the oxygen site, $\chi_s(T)$, from
\begin{equation}
\label{shift}
K (T) = K_{orb} + a \chi_s (T) \ ,
\end{equation}
where $K_{orb}$ is orbital shift due to bound electrons and $a$ is the hyperfine coupling. By lowering the temperature, we observe a reduction in spin susceptibility (Fig.~\ref{fig:shift}), which is a characteristic signature of the opening of a spin gap. \cite{Takigawa_91,Walstedt} Interestingly, in the near-optimal sample ($y = 7.105$) pseodogap still remains substantially open (as can be seen in inset of Fig.~\ref{fig:shift}).

In order to determine the pseudogap characteristic temperature $T^*$ from the temperature dependence of the Knight shift, we have employed an often-used phenomenological three-parameter function: \cite{Fujita,Stern,Mehring_92}
\begin{equation}
\label{KT}
K (T) = \frac{ a \chi_0 }{\cosh^2 \left( T^* / 2 T \right) } + K_{orb} \ ,
\end{equation}
where $K_{orb}$, $a \chi_0$ and $T^*$ are fit parameters. As can be seen from Fig.~\ref{fig:shift}, the fit is very good above $T_c$. 

\section{DISCUSSION}
In Fig.~\ref{fig:pd}a we show our Knight shift results for the $x = 0.1$ and $x = 0.4$ families and compare them with previous CLBLCO $T^*$ data on the same family obtained by Ca--NMR Knight shift measurements \cite{Marchand} and SQUID magnetization. \cite{Lubashevsky_08} As can be seen, this study is in agreement with earlier analysis, with the largest discrepancies arising from SQUID $x$ = 0.1 data. Since NMR directly probes local spin susceptibility on oxygen sites \emph{in the planes}, and the oxygen Knight shift being sensitive only to pseudogap opening (with no large additional signal, as in SQUID measurements), we concur that our $^{17}$O-NMR measurements are the most reliable of the three experimental methods. The general behaviour of the pseudogap onset temperature agrees with other cuprates: $T^*$ is higher for more underdoped samples, and  decreases seemingly linearly with increased oxygen content ($y$). However, the nearly optimally doped samples still show a strong temperature dependence of the spin susceptibility. Also, the family with higher maximum $T_c$ ($x = 0.4$) has lower pseudogap temperatures. This result implies different PG behaviour with respect to SC dome in various families. 

A more uniform picture emerges if we allow for the fact that the oxygen doping $y$ is not a good measure of mobile holes in the planes. As discussed earlier, \cite{Keren_03,Keren_06,Lubashevsky_08,Ofer_08} $\mu$SR and NQR measurements indicate that the proper doping parameter for CLBLCO is $\Delta p$, the difference of in-plane hole doping, $p$, from optimal doping $p_{opt}$ for the given family. $\Delta p$ is related to chemical oxygen doping $y$ via
\[ \Delta p = K(x) (y - y_{opt}) \ , \]
where $y_{opt}$ is the chemical doping of optimal sample and $K(x)$ is a family-dependent parameter that accounts for plane doping efficiency. \cite{Keren_06} If $\Delta p$ is used in place of $y$, all SC domes acquire a similar shape. In Fig.~\ref{fig:pd} we show the scaled phase diagram with results from all families. Although the lack of samples with wide $y$ variation in $x = 0.2-0.3$ prevents us from making any strong assertions there, it can still be seen that all the data is consistent: spectral widths, frequency and temperature behaviour differ only slightly between families. All $T^*$ points collapse roughly onto the same line. 

From the unscaled raw data it is apparent that the family with highest $T_c$ tends to have the lowest $T^*$, and vice versa. Once the `true' hole doping is introduced, the anticorrelation is less pronounced, but it remains that $T^*$ is much less affected by a change in $x$ than the superconducting transition temperature. A weak correlation between the N\'{e}el temperatures of underdoped CLBLCO and $T^{*}$ was observed in a magnetic susceptibility study, indicating a magnetic origin of the pseudogap. \cite{Lubashevsky_08} The intrinsic uncertainty of $T^{*}$, the small variation of $T_{Neel}$ between families and the relatively small number of differently doped samples investigated prevent us from giving definite conclusions about the scaling of $T^{*}$ with $T_{Neel}$ in this study. However, as mentioned above, we do observe that the relative variation of $T^{*}$ with $x$ is much smaller than the change in $T_c$ (noting that the influence of $x$ on $T_c$ is an interesting subject itself \cite{Amit_10}). In view of the relative robustness of $T^{*}$, the prospect of pseudogap as a generic consequence of Mott physics seems appealing to us, \cite{Sordi_12} but detailed calculations and more extensive experiments are needed in order to substantiate the proposition.

The near-constancy of $T^*$ with changing $x$ places constraints on any theory of the pseudogap. If the pseudogap is due to a distinct ordering (such as incommensurate spin density wave, orbital density wave, or similar\cite{Nayak_00,Laughlin_14}), the order must be insensitive to the lattice deformations (and corresponding variation of electronic hopping integrals) brought upon by changing $x$. \cite{Keren_06,Ofer_08} Regarding the position of the tentative quantum critical point in relation to the superconducting dome, it is obvious that \emph{in this system} it does not lie near optimal doping, since here $T^*$ is close to 200~K. Although we cannot reliably determine the position of the QCP from available data, it is clear that it is inconsistent with the proposed location close to optimal doping. \cite{Varma_99} Also, we observe a significant difference from experimental results on the similar YBCO, \cite{Fauque_06,Kapitulnik_09} where the QCP appears to nearly coincide with maximum $T_c$. From our results it seems that position of quantum critical point is not universally close to optimum doping, but depends on the details of the system. In that case high-temperature superconductivity cannot be simply related to quantum-critical fluctuations.

\section{SUMMARY}
To conclude, we have detected a spin gap in the Knight shift of in-plane oxygen atoms in the cuprate family CLBLCO, confirming the universality of the pseudogap in cuprates and gaining insight into the nature of the pseudogap state. Our systematic $^{17}$O investigation of CLBLCO families shows that the local level of chemical disorder is similar in all investigated samples, making them suitable for investigating purely electronic effects. We observe different behaviour of pseudogap and superconductivity among different CLBLCO families, indicating that pseudogap and superconductivity are not directly correlated. Our data also indicate that the pseudogap extends into the overdoped side of the phase diagram, in contrast to results on the well-known, structurally similar cuprate YBCO. This suggests that the relation between pseudogap line and superconducting dome observed in YBCO is not universal for the cuprates. As the in-plane electronic parameters vary smoothly across all investigated CLBLCO families, the interplay between superconductivity and pseudogap is here visible with unprecedented clarity, providing an important benchmark for theories of the pseudogap state.

\section*{ACKNOWLEDGMENTS}
We gratefully acknowledge helpful discussions with M. S. Grbi\'{c} and A. Dul\v{c}i\'{c}.
The research leading to these results was supported by equipment financed from the European Community (FP7) under grant agreement no.\ 229390 SOLeNeMaR and by the Israeli Science Foundation (ISF). TC, DP, and MP acknowledge support by the Croatian Science Foundation (HRZZ) under grant no.\ 2729.

\bibliography{CLBLCO_lit}
\end{document}